\documentclass[twocolumn]{aastex631}

\usepackage{ulem}

\begin{document}

\title{}

\correspondingauthor{Meridith Joyce}
\email{mjoyce@stsci.edu}

\author[0000-0002-8717-127X]{Meridith Joyce}
\affiliation{Space Telescope Science Institute,
3700 San Martin Drive,
Baltimore, MD 21218, USA}
\affiliation{Kavli Institute for Theoretical Physics, University of California, Santa Barbara, CA 93106, USA}
\altaffiliation{Lasker Fellow}

\author[0000-0002-4818-7885]{Jamie Tayar}
\affiliation{Department of Astronomy, University of Florida, Bryant Space Science Center, Stadium Road, Gainesville, FL 32611, USA }
\affiliation{Kavli Institute for Theoretical Physics, University of California, Santa Barbara, CA 93106, USA}

\author[0000-0002-7635-9728]{Daniel Lecoanet}
\affiliation{Department of Engineering Sciences and Applied Mathematics, Northwestern University, Evanston IL 60208, USA}
\affiliation{CIERA, Northwestern University, Evanston IL 60201, USA}
\affiliation{Kavli Institute for Theoretical Physics, University of California, Santa Barbara, CA 93106, USA}

\title{Gender Disparity in Publishing Six Months after the KITP Workshop \textit{Probes of Transport in Stars} }

\begin{abstract}
Conferences and workshops shape scientific discourse.
The Kavli Institute for Theoretical Physics (KITP) hosts long-term workshops to stimulate scientific collaboration that would not otherwise have taken place.
One goal of KITP programs is to increase diversity in the next generation of scientists.
In this analysis, we examine gender trends in authorship of papers generated as a result of the KITP program \textit{Probes of Transport in Stars}, which ran from October 11th, 2021 to December 17th, 2021. While 38\% of workshop participants were women, only 19\% of publications produced between December 1st, 2021 and June 3rd, 2022 had female first-authors. Further, of these early publications, 61\% had all-male author lists. Among publications resulting from the KITP program, the proportions of both male first-author papers and papers with all-male author lists are higher than predicted by models that take into account the gender distribution of the KITP participants. These results motivate more thorough investigations of collaboration networks at scientific conferences and workshops. Importantly, they also suggest that programs, conferences, and workshops of any kind need to take steps beyond those implemented in this KITP program to enable more diverse collaborations and address gender disparities in science. 
\end{abstract}

\keywords{Gender equity in physics and astronomy}

\section{Introduction}
\label{sec:intro}
Many authors have discussed the impact of gender on astronomy careers, including the impacts on time allocation \citep{Reid2014, Patat2016, JohnsonKirk2020, Carpenter2022}, authorship and citation \citep{Caplar2017}, hiring \citep{Flaherty2018, Perley2019, Kewley2021}, conference participation \citep{Davenport2014, Pritchard2014, Schmidt2016, SchmidtDavenport2017}, and involvement in collaborations \citep{Lundgren2015, LucatelloDiamondStanic2017}.

A method often employed in other fields is the study of authorship networks, the goal of which is understanding with whom people publish. In turn, this can serve as an imperfect indicator of their choice of collaborators \citep{KatzMartin1997}. 

There has been significant discussion in the literature on the question of whether authorship networks differ, on average, by gender. Initial results suggested that men have more collaborators, but when controlling for respondent seniority, data suggest that women actually have the same or perhaps even more collaborators than men, on average \citep{BozemanGaughan2011,West2013}. Some authors have found that women are more likely to collaborate with other women \citep{BozemanCorley2004, Karimi2019}, men are more likely to collaborate with men \citep{BozemanGaughan2011, Araujo2017}, or both \citep{FerberTeiman1980,McDowellKilholm1992,Jadidi2018},
but the details of these findings are affected by other variables including career status, field, funding, strategy, and so on \citep[e.g.][]{RhotenPfirman2007}.  

Crucially, the literature suggests that wide, well-connected collaboration networks can be the key to long-term productivity for people of both genders (\citealt{KyvikTeigen1996,Jadidi2018}, although see also \citealt{LeeBozeman2005}). This makes it important to consider not only the intentions of workshops to increase inclusivity and enhance networks, but also their measurable impact on collaboration and publication. 

In the Fall of 2021, the Kavli Institute for Theoretical Physics (KITP) ran a 10-week program on \textit{Probes of Transport in Stars}, as well as a week-long associated conference on \textit{Transport in Stellar Interiors}. There were 45 in-person participants; 28 self-identified as male, and 17 self-identified as female. In the six months after the end of the program, defined as the period of December 1st, 2021 through June 3rd, 2022, the in-person participants in the \textit{Probes of Transport in Stars} program have authored a total of 94 papers/scientific contributions that have been uploaded to the arXiv. Of these scientific contributions (hereafter also ``papers," interchangeably), we have identified 18 as being strongly associated with the KITP program. In this work, we investigate the gender distribution of the authors of these papers who were also program participants. Although our sample size is small, our study is significant in that it has isolated the (early) collaborative effects of a single workshop.

In the roughly six months following the end of the program, men have published disproportionately more than women (Section \ref{sec:results}). We also show that men co-authored papers at a disproportionately higher rate than women. The number of papers with only male authors (amongst program participants) is much higher than if coauthors were chosen at random.
Collectively, these and related findings point to significantly greater publication opportunity for men despite no apparent difference in the resources provided by KITP on the basis of gender (Section \ref{sec:KITP_gender_goals}). Discussion of possible reasons for these biases and output discrepancies is given in Section \ref{sec:conclusions}.
Suggestions for improving gender-based returns from future programs are given in Section \ref{sec:improvement}.

\section{Gender Goals of KITP Programs}
\label{sec:KITP_gender_goals}
As part of its mission to empower the next generation of scientists and foster collaborative networks (https://www.kitp.ucsb.edu/support-kitp), KITP requires program organizers to consider inclusivity seriously in the creation of a program. At the pre-proposal stage, organizers are required to suggest ``2-4 organizers and 10-15 key participants, keeping in mind KITP's commitment to diversity and inclusion of under-represented groups.'' At the full proposal stage, KITP works with the organizers to ``identify and attract a diverse set of scientists through an application and invitation process,'' and organizers are required to name a Diversity Coordinator as well as a ``diverse list of Key Long Term Participants.'' 

As recruitment begins, organizers are reminded that the program should be well advertised in order to attract a broad range of applicants from which a diverse pool of participants can be chosen. They are also encouraged to ensure that speakers and session chairs at the associated conference should be broadly diverse along a range of axes, including gender, race, under-representation, career stage, and so on. 
Before finalizing the program, KITP required a short report on these recruitment efforts, including demographic information on the proposed participants. In building the list of conference speakers, organizers were reminded that they ``must consider the inclusion of groups underrepresented in physics.'' At the close of the program, coordinators were expected to do an exit interview as well as produce a report on the program itself, which included a review of ``efforts at achieving participation by groups underrepresented in physics.'' KITP itself has also begun collecting data on collaboration networks and the impact of programs, which will likely be extremely valuable and potentially influential in setting future policies.

\subsection{The ``Probes of Transport in Stars'' Program}
A specific goal of the \textit{Probes of Transport in Stars} program was to encourage new collaborations, particularly between participants who are interested in similar astrophysical problems but normally use different approaches (3D simulations, 1D models, observations, analytical calculations, etc).
During the participant selection process, the organizers kept statistics on the methodological expertise of the applicants, as well as demographic information including whether an applicant was likely part of an underrepresented minority.
Ultimately, the constituency of in-person participants was 38\% female. This is significantly better than the degree of female representation in professional astronomy (27\% female) and even more so than in professional physics (20\% female).\footnote{according to the American Institute of Physics (AIP) records on gender (2019--2020): 
https://www.aip.org/statistics/data-graphics/percent-astronomy-faculty-members-who-are-women-2003-2020}

A variety of mechanisms were put in place to encourage inter-speciality cross-talk and the creation of new projects:
\begin{enumerate}
\item Each week, the in-person schedule included an introduction and organization session on Monday mornings;
\item Several broad working groups were created that met once a week;
\item Formal talks to introduce various concepts and techniques happened twice a week;
\item Informal meetings on Fridays at which each participant was encouraged to bring and briefly present a plot representing their efforts over the week;
\item Participants were invited to group lunches many weekdays; 
\item Group dinners were organized on Tuesdays and Fridays; and
\item A variety of social activities were organized and advertised by participants. 
\end{enumerate}
In addition to the in-person events, a KITP Slack workspace was open to all participants, and a set of Google Drive folders was created for the storage of working group notes and project efforts. Working group chairs and project leaders were encouraged to create a Slack channel for their project that included a description and summary of current work and to make these channels open to any participant (including remote participants) interested in joining. These channels were intended to permit collection and fragmentation as group interests dictated.  

The week-long conference \textit{Transport in Stellar Interiors} was organized to focus on discussions. Each three-hour session was divided into a 40-minute review talk (plus 20 minutes of questions), followed by a 20-minute research talk (plus 10 for questions), followed then by a 90-minute panel discussion. The organizers prioritized junior researchers with a range of identities as speakers and panelists. During the conference itself, session chairs were reminded to be inclusive when mediating conversation and were provided with guidelines for inclusive chairing.

\section{Data}
\label{sec:data}
\begin{table*} 
\centering 
\begin{tabular}{l l}  
\hline\hline 
Category Label & Definition \\ \hline
\textit{KITP Participant} & an in-person participant in the KITP program for any number of weeks;  \\
 & excludes conference-only participants and remote-only participants \\ \hline
KITP author & a \textit{KITP participant} whose name appears in any position on any paper published from Dec 1, 2021 \\
 & to June 3, 2022 \\ \hline
\textit{KITP Paper} & any paper published by a KITP author whose production is credited to KITP (per Section \ref{sec:data}) \\ \hline
\textit{all-female} &   \\
\textit{KITP paper} & any \textit{KITP paper} for which all \textit{KITP authors}$^{\star}$ are female \\ \hline
\textit{all-male} &    \\
\textit{KITP paper} & any \textit{KITP paper} for which all \textit{KITP authors}$^{\star}$ are male  \\ \hline
\end{tabular}
\caption{Publications by arXiv ID. $^\star$Note that such author lists may contain men (women), but those men (women) are not \textit{KITP authors} by definition. As such, this is a measure of clustering among female/male \textit{KITP participants}.}
\label{table:definitions}
\end{table*}

For this analysis, ``participation'' in the KITP workshop refers only to in-person participation. By this metric, and excluding those who attended only the mid-program conference, there were 45 participating scientists in total.\footnote{The in-person participants list is available here: \url{https://online.kitp.ucsb.edu/online/transtar21/directory.html}} 

The data set was constructed first by performing a NASA ADS search on each of the 45 in-person participants over the date range December 1, 2021 to June 3, 2022. For an author named $n$, any paper on which $n$ appears, in any position, is classified as ``a paper by $n$.'' The total number of papers returned this way is 94, and the arXiv IDs for each of these scientific contributions are given in Table \ref{table:papers} in the Appendix. 

Sub-classifying a paper published between December 1st, 2021 and June 3rd, 2022 by author $n$ as a \textit{KITP paper} by author $n$ is attempted as follows. First, we define a ``KITP paper'' as any paper or other arXiv/ADS-supported scientific contribution whose inception or production began in one of three ways:
\begin{enumerate}
    \item as a new idea while a portion of its authors were at KITP (and not before); or
    \item as an iteration of an existing idea or goal whose actualization was only possible thanks to work at KITP; or  
    \item after KITP, but as a direct consequence of work performed at KITP, where such work includes the creation of collaboration networks.
\end{enumerate}
One means of verifying a paper's membership in the \textit{KITP paper} category is checking whether that paper includes the following grant acknowledgement or a variant thereof: \texttt{This research was supported in part by the National Science Foundation under Grant No. NSF PHY-1748958.} However, this is not a perfect indicator for a variety of reasons: among these, (1) whether citing this grant is justified or necessary is up to the author's discretion; (2) its citation, or lack of citation, may not reflect true working conditions for political or other ambiguous reasons; or (3) authors may simply have forgotten to cite KITP. We thus found that a qualitative method was most effective for determining whether work was enabled by the KITP program.

\begin{figure*}
    \centering
    \includegraphics[width=\textwidth]{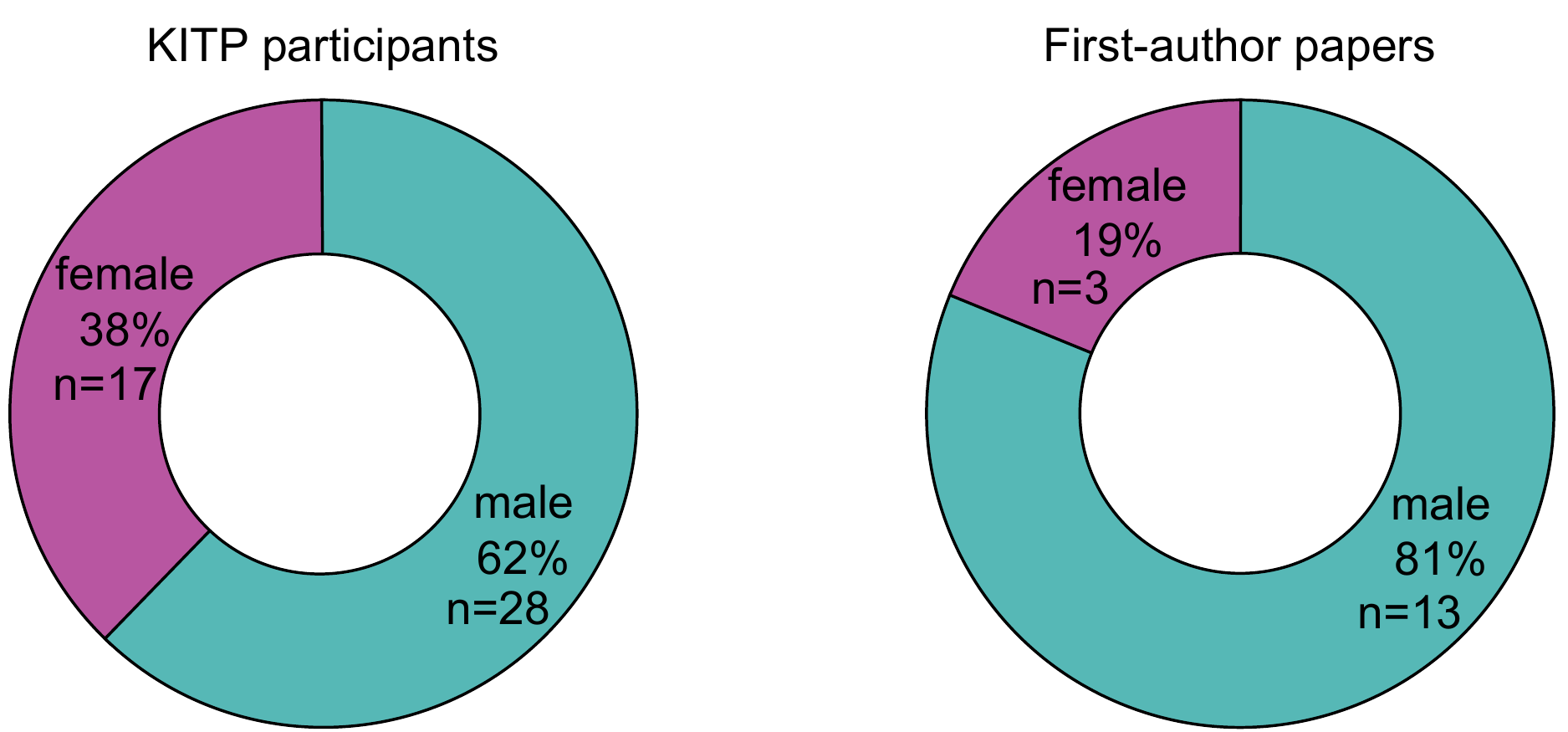}
    \caption{\textit{Left:} Gender breakdown of KITP program participants. \textit{Right:} Gender breakdown of the first-authors of KITP papers. We note that the first-authors of two KITP papers were not KITP participants, and so they are not included in the right-hand panel. 
    }
    \label{fig:piecharts1}
\end{figure*}

The definition of ``arXiv/ADS-supported scientific contribution'' (or ``paper'') extends to (RNAAS) notes, publication pre-prints (at any stage of refereeing, including unrefereed), novelty/recreational papers, and manuscripts not intended for publication in astronomy or physics journals. Conference proceedings were included where they were substantial enough to make it through arXiv moderation. 

The number of papers identified as KITP papers according to these criteria is 18. Of these, 16 are led by a KITP author, which we define as an (in-person) program participant who appears somewhere on at least one paper published between December 1, 2021 and June 3, 2022. There is a potential maximum of 45 such authors, but in practice, fewer. 

Many KITP papers have authors (including first authors) who are not KITP participants. In the subsequent analysis, we only study the gender breakdown of the \textit{KITP participants} who were authors of these papers. We have self-identified genders for the KITP participants, but we do not have this information for other (non-KITP) authors on the papers. Furthermore, the goal of this work is to learn about collaboration patterns between KITP participants, specifically.

The categorizing definitions used in this analysis are summarized in Table \ref{table:definitions}. 

\subsection{Caveats}
\label{sec:caveats}
Social data are even more imperfect than astronomical data, and so our classification scheme comes with many caveats:

\begin{enumerate}
    \item  ``KITP paper'' is not a rigid definition. It is not straightforward to determine whether a piece of work has come to fruition entirely, partially, or not at all due to KITP conditions. This is a subjective assessment that would, ideally, be made by each author individually for each paper on which they appear. However, it is not possible to collect these data on a reasonable timescale and without biasing our results.
    Where we have classified a paper as a \textit{KITP paper}, we have done so using our best judgement: factoring in the author list, time of publication, topic of publication, presence of grant acknowledgement, and firsthand knowledge of working groups, subject areas, and projects in progress during the workshop.
   
    \item The arXiv does not capture works in progress, for example, works submitted to journals but still in revision or under consideration. The arXiv also cannot capture intangible benefits of collaboration, such as the formation of new (potential) author networks or the initialization of longer-term projects for which there is no searchable data product (yet). 
    
    \item Genders of the KITP authors and participants are taken to be those self-reported by those individuals. As part of KITP registration, individuals were given the option to choose from among \verb|woman|, \verb|man|, \verb|non-binary|, or ``\verb|another identity not listed|'' as identifiers. Use of the terms men/women and male/female throughout this manuscript shall not be taken to imply a lack of other gender identities, within this sample or outside of it. 
\end{enumerate}

\section{Results}
\label{sec:results}

\begin{figure*}
    \centering
    \includegraphics[width=\textwidth]{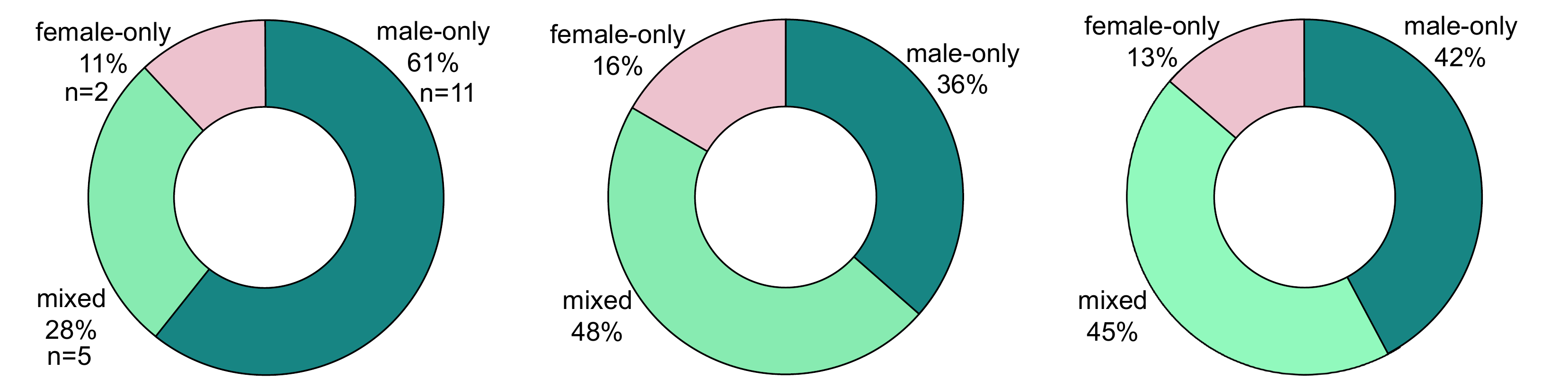}
    \caption{\textit{Left:}
    Observed gender breakdown of KITP papers.
    \textit{Center:} ``Random Authors'' model: the expected gender breakdown of authors of KITP papers, drawing from the 62-38 M:F ratio of the workshop. 
    \textit{Right:} ``Random coauthors'' model: the expected gender breakdown of the authors of KITP papers, using the observed gender breakdown of first authors, and drawing from the 62:38 M:F ratio of coauthors. 
    Note that we only include the gender of KITP participants (for whom we have have self-reported genders). Some ``female-only'' papers may include non--KITP-participant male authors, and some ``male-only'' papers may include non--KITP-participant female authors. The presence of additional (non-KITP) authors does not affect our assessment of the gender-based clustering within the KITP program.
    }
    \label{fig:piecharts2}
\end{figure*}

We summarize our quantitative findings below. Interpretation is reserved for Section \ref{sec:conclusions}. 

We discuss our findings in the context of an ``idealized environment'' from which random gender selections are made. Such an environment assumes that women and men have no preference for or against their own gender when choosing collaborators and are exactly the same in terms of their ability to publish, opportunity to publish, access to the resources to publish, motivation to publish, and preference for publication style and type. The KITP participants were 38\% female and 62\% male (Figure~\ref{fig:piecharts1}). In an environment as described, we would expect the gender breakdown of KITP first authors to reproduce the participant ratio. In reality, of 16 KITP first authors (excluding two KITP papers which do not have a KITP participant as first author), only 3 had female first-authors, while 13 had male first-authors. This is half the female first-author representation idealized conditions would be expected to produce. 

In the left-most panel of Figure \ref{fig:piecharts2}, we show the actual distribution of female-only, male-only, and mixed-gender KITP papers observed in the data. In the center and right-most panels, we show results from simulated distributions computed two ways. 
The center panel shows the expected values for numbers of female-only, male-only, and mixed-gender papers assuming a ``Random Authors'' model. In this model, the genders of authors for 18 synthetic papers are randomly drawn from a 38:62 F:M distribution, preserving the observed number of authors associated to each paper. In the right-most panel, we assume a ``Random Coauthors'' model. In this case, the reported genders of the first authors (Figure~\ref{fig:piecharts1}) are fixed. Then, each remaining coauthor's gender is randomly drawn as in the Random Authors model. This model yields somewhat, but not wildly, different expected values for the numbers of single-gender papers.

We first note that the percentage of KITP papers containing at least one female KITP author is 39\%,
compared to an expected value of 64\% from the Random Authors model or 58\% from the Random Coauthors model. This shows that the actual proportion of KITP papers containing at least one female KITP author is significantly lower than what would be produced via random draw, even if the observed proportion of male first-authors were the same. The percentage of KITP papers containing at least one male KITP author, on the other hand, is 89\%. This is larger than the percentage of male KITP participants (which is 62\%), larger than the expected number of papers containing a male author (82\%) according to the Random Authors model, and slightly larger even than the expected number of such papers (87\%) according to the Random Coauthors model. We comment on specific features of the gender distribution data below.

\subsection{Low proportion of female first authors}
Of the 18 KITP papers, 16 have KITP participants as first authors. Only 19\% of these had female first authors, which is lower by a factor of two than the fraction of female participants (Figure~\ref{fig:piecharts1}). Out of 16 papers, a random draw would produce 6 female-led KITP papers. We instead see three, and in all cases, the number of coauthors is either one or zero (in fact, this is true whether we consider KITP coauthors or coauthors of any kind). This is in stark contrast to male-led papers, which range in number of KITP coauthors from zero to five (and up to dozens if including non-KITP coauthors). 

\subsection{Men prefer male collaborators}
\begin{figure}
    \centering
    \includegraphics{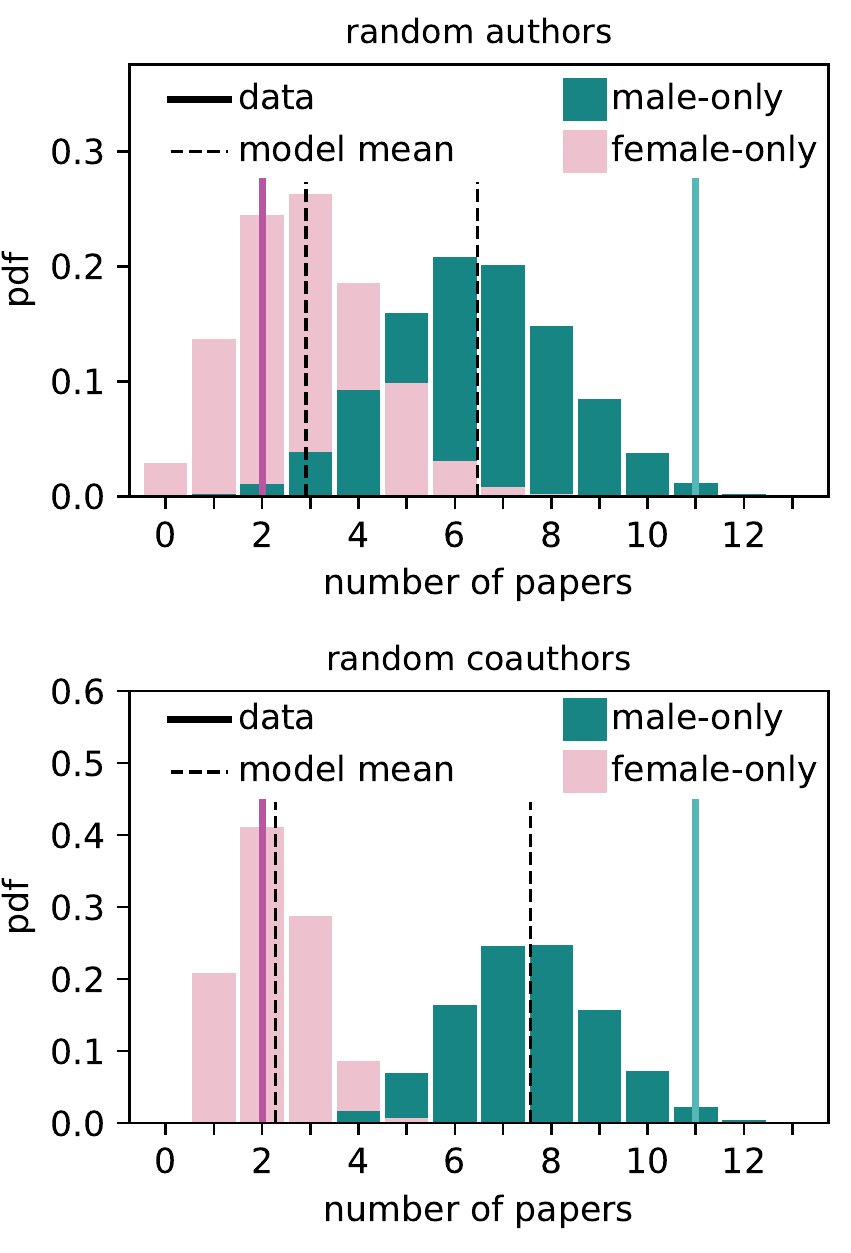}
    \caption{Probability distributions for the expected number of male-only and female-only papers assuming the Random Authors (top) and Random Coauthors (bottom) models are shown.
    Under the constraints of the former model, the expected number of female-only papers is 2.91 (vertical dashed line on the pink distribution), which over-predicts the observed number of female-only papers (solid pink line) by one (discretized). The expected number of all-male KITP papers, however, is 6.46 (vertical dashed line on the blue distribution), which is 5 papers (discretized) less than observed (solid blue line).
    Under the constraints of the latter model, the expected number of female-only papers is 2.28, and the expected number of male-only papers is 7.57 (vertical dashed lines). In this case, the expected number of female-only papers (2.28) closely matches the observed number of female-only papers (2, solid line), but the expected number of male-only papers (7.57) is still very different from the observed value (11, solid line).}
    \label{fig:histograms}
\end{figure}

\textbf{Most KITP papers have all-male KITP author lists.}
Per Figure~\ref{fig:piecharts2}, of the 18 KITP papers, 11 have author lists with only male KITP participants (\textit{all-male KITP papers}), whereas two have authors lists with only female KITP participants (\textit{all-female KITP papers}). This constitutes a large over-representation of all-male papers even when taking into account the larger fraction of male first authors.

Figure~\ref{fig:piecharts2} also makes clear that the degree of single-gender author clustering is much higher in the observed data than predicted by either model. The proportion of all-male KITP papers (61\%) alone exceeds the predicted number of mixed-gender papers by either model. When comparing to the distribution generated by the Random Coauthors model, the true number of all-male KITP papers (61\%) exceeds both the predicted number of all-male KITP papers (42\%) and the expected number of papers that contain a female author in any position at all (58\%). 
It is thus clear that male authorship is over-represented, by any measure, in the real data. 

There are not enough data in the category of female-led publications to support inferences about women's gender preferences among coauthors. However, the data on male-led papers demonstrate a clear preference among male leads for male coauthors.

To better understand the degree of over-representation and quantify how anomalous it is, we computed probability distributions for the numbers of male-only and female-only KITP papers under the assumptions of both the Random Authors and Random Coauthors models. Results are shown in Figure~\ref{fig:histograms}.

In the probability distribution built on the Random Authors model, the expected number of all-male KITP papers is 6.46 and the expected number of all-female papers is 2.91, indicated by dashed vertical lines in the centers of the respective distributions. The observed values are indicated as pink/blue solid lines for all-female and all-male, respectively. These lie at the far extremes of the probability distributions in opposite directions. In the probability distribution built on the Random Coauthors model, the mean number of male-only papers is 7.57 and the mean number of female-only papers is 2.28. While the actual number of all-female papers (2) is near the peak of this distribution, there are still many more all-male papers (11) than would be expected if KITP coauthors were chosen at random. The probability of observing 11 or more male-only papers given these assumptions is $p=0.03$. Thus, even taking into account the disproportionately large number of male first-authors of KITP papers, there is a demonstrable bias among male KITP authors in favor of collaborating with male KITP participants.

On the other hand, the true number of female-only papers is nearly the same as predicted by either model: two papers as compared to the (discretized) expected values of three papers (Random Authors) and two papers (Random Coauthors). As such, the models do indicate that the majority of single-gender clustering overall is the result of all-male groups. However, it is worth noting that two-thirds of (2 out of 3) KITP papers with female first-authors are female-only. This is higher than the same ratio for men, suggesting that female preference for female coauthors may also be a relevant dynamic. That said, it is difficult to draw meaningful conclusions from samples of this size, so we can neither confirm nor rule out single-gender clustering by female authors.

\subsection{More publication opportunities for men}
\begin{figure*}
\begin{center}
\includegraphics[width=\textwidth]{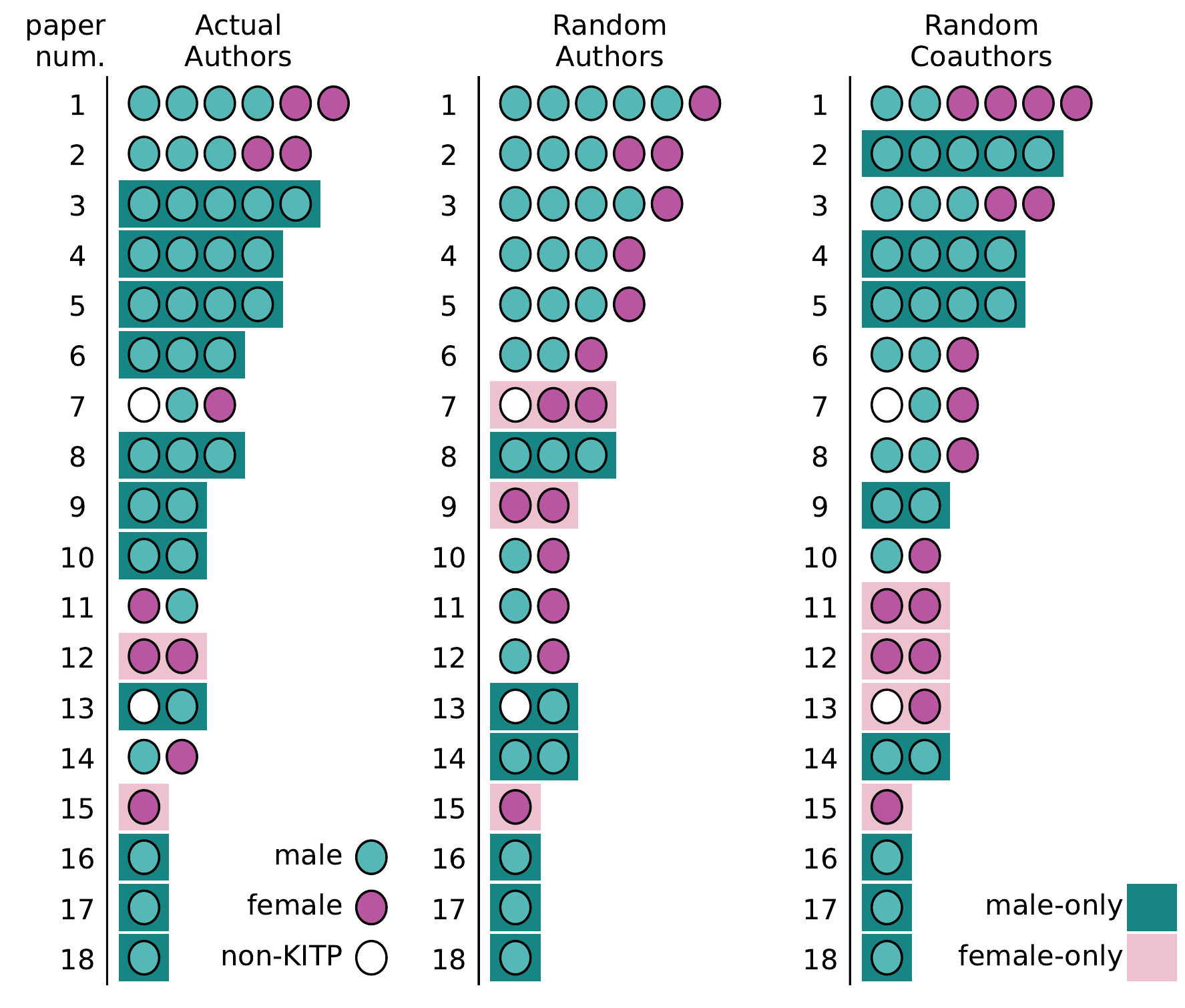}
\caption{ 
\textit{Left:} The observed gender breakdown of authors on KITP papers. 
\textit{Center:} One realization of the expected gender breakdown of authors on KITP papers if the KITP authors on each paper were drawn randomly from the participants.
\textit{Right:} One realization of the expected gender breakdown of coauthors on each paper were drawn randomly from the participants, with the observed genders of first authors fixed.
Blue circles indicate male authors, pink circles indicate female authors, blue bars indicate all-male KITP papers, pink bars indicate all-female KITP papers, and the two white circles that appear in every column represent the two KITP papers not led by a KITP author. We note again that any ``all-male'' or ``all-female'' KITP paper may have non-KITP authors of any gender; we assess KITP clustering only. 
}
\label{fig:gender_tree}
\end{center}
\end{figure*}
Figure~\ref{fig:gender_tree} is an extension of Figures~\ref{fig:piecharts2} and \ref{fig:histograms}, showing the gender breakdown of KITP author lists according to number of KITP authors per paper. The true distribution (left column) is compared to synthetic distributions from the Random Authors (middle column) and Random Coauthors (right column) models. The indices align a given KITP paper to its simulated counterparts per row (and do not necessarily correspond to the indices assigned to arXiv identifiers in the Appendix). 

Figure~\ref{fig:gender_tree} provides striking visual indication of how (early) publication opportunities associated with the KITP program were distributed by gender: men preferentially selected male collaborators, published more first author papers, and disproportionately did not include women on their papers. Both models predict that the number of mixed-gender papers should exceed the number of all-male papers, but the data do not reflect this. The over-representation of all-male KITP papers found in the data is especially noticeable on papers with three or more coauthors, suggesting that the preference among men for male collaborators may be compounded as number of male authors increases. 

\vspace{1cm}
\section{Conclusions \& Discussion}
\label{sec:conclusions}
Despite considerable effort on behalf of the \textit{Probes of Transport in Stars} organizers (two of whom are authors on this manuscript) and the Kavli Institute for Theoretical Physics itself (see Section~\ref{sec:KITP_gender_goals}), neither equitable nor proportionate gender outcomes were achieved according to the metric of publishing. Even allowing for an over-representation of male first authors---which may occur naturally for a variety of reasons, including that men are over-represented in more advanced career stages and/or that men have, on average, less competition for their mental resources---the number of all-male KITP papers is anomalously high. This result is very unlikely to have been observed in conditions that reflect, or mostly reflect, the assumptions of the ideal collaboration environment described in Section~\ref{sec:results}.

The presence of all-male clustering
is strongly supported by the low likelihood ($p<0.03$) of producing 11 male-only KITP papers given the program's gender distribution (Figure~\ref{fig:histograms}). While the number of female-led papers is too small to make quantitative arguments regarding, for example, bias against women in lead authorship roles, the facts that (1) there are so few such papers, and (2) that the maximum number of coauthors on these papers is 1, are not inconsistent with a preference against female collaborators among men.

We suspect that the most likely cause of the over-representation of male authors is a preference among men for male collaborators. Other explanations are also possible, including those that intersect with or amplify the effects of male preference for male collaborators. For instance, women may have produced proportionally fewer lead-author publications due to unequal demands on their time: they may have had a greater need to devote their time at KITP to finishing existing projects, or perhaps they could not escape from non-work-related burdens and time constraints (e.g., childcare, pet care, household management) to the same extent as their male peers. It is also possible that women were disproportionately excluded---whether consciously or unconsciously by men, or by their own choice---from networking or social events. As social events in academic spaces are, by and large, \textit{de facto} collaboration opportunities, one could imagine that self-segregation by gender socially could readily propagate into self-segregation by gender on academic projects. This social self-segregation could be particularly influential for determining coauthors, as it is not always necessary to make a significant contribution to a paper to be invited as a coauthor. 

Yet another possibility is that there are preferences by gender regarding publication type. These data, measured at a relatively early time post-program, could be explained by, e.g., a preference among women for longer-term projects or a preference among men for faster publication timelines. We have not studied these possibilities here, as they require more longitudinal data, but future analysis may shed light on these hypotheses.

Though KITP provided a nearly obligation-free \textit{scientific} and \textit{working} environment to all, broader gender inequalities regarding free time, expected labor, mental availability, and social segregation may still have exerted influence. To mitigate the impacts of these and other well-documented productivity disadvantages affecting women requires thoughtful policy beyond ``equal opportunity'' or representative gender distributions on participants lists. Some actionable suggestions are given in the next Section.

\section{Suggestions for Improvement}
\label{sec:improvement}
It is often observed anecdotally that women and gender minorities are aware of gender bias and men are not. Regardless of the degree to which this is true in practice, the first step towards equitable outcomes is awareness of a lack of equality despite efforts to encourage it.

First, it is imperative that we re-assess gender outcomes from this workshop at later dates and regular intervals. We suggest that the publication data sets should be reconstructed again in 6 months (December 2022) and 12 months (June 2023) to see how trends evolve. It is especially important to learn whether the striking gender disparity in research output we see now lessens over time, though this will not affect conclusions drawn now, at the 6-month stage. 

We must also recognize that the strategies employed by KITP and the program's organizers were insufficient for preventing a high degree of gender clustering, especially among men. To this end, we offer some ideas for discouraging gender clustering, improving participant cross-talk, and avoiding the pitfalls of social dynamics that, intentionally or unintentionally, make women less likely to be included as collaborators. Though a specific KITP program was the context for this study, these suggestions apply to all programs and workshops.

\begin{itemize}

\item[] Consider opening your program or workshop with a lecture on gender statistics to ensure that those who are not already aware of gender bias in publishing and collaborating are made aware, and consider reiterating this information at, for example, the start of each week 

\item[] Consider giving a short assessment to potential participants to learn whether they are aware of gender bias 

\item[] Consider how opportunities for self-segregation by gender could be limited; for example, by introducing structure around mealtimes

\item[] Consider whether social events designed for academic purposes are taking place in gendered spaces,\footnote{spaces where members of one gender may feel more comfortable than members of another, or spaces where it is safer for one gender than another} such as bars, and whether collaboration opportunities are being scheduled at times where women proportionally have more home obligations (e.g. after standard working hours, on school nights, etc.)

\item[] Consider incentivizing multi-gender collaboration with actions such as verbally surveying who worked with whom over a given time interval or otherwise measuring, monitoring, and drawing attention to the gender composition of breakout working groups throughout the program

\item[] Work to ensure that men take on the same or greater amounts of organizational labor and non-science-related burdens while participating in a program; social science data show that women will tend to volunteer themselves for these obligations out of social conditioning, so it is best to assign non-science-related labor externally

\item[] Strive to model gender inclusivity by publicizing inclusion as a ``best practice'' of the program environment and of science more broadly; for example, by stating that inclusive behavior is not only encouraged, but expected in your program

\item[] Consider devising a formal system for declaring and joining projects made possible by your program's resources, as is the norm, for example, in large observational collaborations

\item[] Consider distributing reports like this one to participants prior to the program 

\end{itemize}

As physicists and astronomers, we are not trained, in a professional sense, to study gender bias. We have therefore tried to limit the complexity and causal analysis of our study to the the most clear conclusions, namely that men are over-represented among first authors and that the data support a preference among men for male coauthors. Both of these conclusions are consistent with a variety of previous studies in a wide range of fields, from formal social science to other demographic studies within physics. However, we emphasize them here in the practice of modern astronomy, in hopes that our study can encourage future efforts to improve gender discrepancies in astronomy careers. While we do not expect the suggested mitigation strategies, even if perfectly executed, to guarantee equitable gender outcomes, we look forward to continued discussion and efforts to improve the state of our profession.

\begin{acknowledgments}
M Joyce wishes to thank John Bourke for tolerating statistical discussions. M Joyce acknowledges the Lasker Data Science Fellowship provided by the Space Telescope Science Institute. 
D Lecoanet wishes to thank Jia Luo for aid in conceptualization and data visualization.
M Joyce also wishes to thank KITP for inviting her to this program regardless of the fact that, on some level, they must have known she was going to write this paper.
This research has made use of NASA's Astrophysics Data System Bibliographic Services.
This research was supported in part by the National Science Foundation under Grant No. NSF PHY-1748958.
\end{acknowledgments}

\software{NASA ADS, arXiv} 

% \newpage
\appendix 
\section{Papers included in this analysis}
We list each member of the data set on which the above statistics were calculated by arXiv identifier, given in Table \ref{table:papers}. The papers classified as ``KITP papers'' are listed first and total 18. The remaining 76 papers are papers published between December 1, 2021 and June 3, 2022. 

\newpage
\begin{table} 
\centering 
\caption{Publications by arXiv ID} 
\begin{tabular}{|ccc|ccc|ccc|ccc|}  
\hline\hline 
% &	arXiv ID &	KITP paper? \\ \hline 
index	&	arXiv ID 	&	KITP?	&
index	&	arXiv ID 	&	KITP?	&
index	&	arXiv ID 	&	KITP?	&
index	&	arXiv ID 	&	KITP?	\\\hline
1	&	2206.00011	&	yes	&
25	&	2205.09655	&	no	&
49	&	2204.00661	&	no	&
73	&	2112.02026	&	no	\\
2	&	2205.02251	&	yes	&
26	&	2205.08841	&	no	&
50	&	2203.15365	&	no	&
74	&	2201.04140	&	no	\\
3	&	2202.10026	&	yes	&
27	&	2201.01722	&	no	&
51	&	2203.14538	&	no	&
75	&	2110.11974	&	no	\\
4	&	2205.09903	&	yes	&
28	&	2205.09125	&	no	&
52	&	2203.15116	&	no	&
76	&	2201.00891	&	no	\\
5	&	2205.03319	&	yes	&
29	&	2205.07964	&	no	&
53	&	2109.13840	&	no	&
77	&	2112.12800	&	no	\\
6	&	2204.10875	&	yes	&
30	&	2205.07996	&	no	&
54	&	2203.11532	&	no	&
78	&	2112.12122	&	no	\\
7	&	2204.08487	&	yes	&
31	&	2202.12902	&	no	&
55	&	2203.11809	&	no	&
79	&	2111.11552	&	no	\\
8	&	2204.00002	&	yes	&
32	&	2202.07524	&	no	&
56	&	2203.11227	&	no	&
80	&	2111.14047	&	no	\\
9	&	2203.11071	&	yes	&
33	&	2203.08920	&	no	&
57	&	2202.10507	&	no	&
81	&	2112.05964	&	no	\\
10	&	2203.09525	&	yes	&
34	&	2205.03020	&	no	&
58	&	2110.11356	&	no	&
82	&	2111.11633	&	no	\\
11	&	2203.06186	&	yes	&
35	&	2205.01860	&	no	&
59	&	2203.05463	&	no	&
83	&	2110.06220	&	no	\\
12	&	2203.06187	&	yes	&
36	&	2205.02278	&	no	&
60	&	2203.04970	&	no	&
84	&	2110.01565	&	no	\\
13	&	2203.02046	&	yes	&
37	&	2205.01798	&	no	&
61	&	2110.03261	&	no	&
85	&	2111.06434	&	no	\\
14	&	2202.03440	&	yes	&
38	&	2205.01298	&	no	&
62	&	2202.12903	&	no	&
86	&	2111.06891	&	no	\\
15	&	2111.01958	&	yes	&
39	&	2204.12643	&	no	&
63	&	2202.11080	&	no	&
87	&	2112.03306	&	no	\\
16	&	2111.01959	&	yes	&
40	&	2202.02373	&	no	&
64	&	2202.08398	&	no	&
88	&	2112.01309	&	no	\\
17	&	2201.10567	&	yes	&
41	&	2204.09739	&	no	&
65	&	2202.07811	&	no	&
89	&	1509.03630	&	no	\\
18	&	2201.10519	&	yes	&
42	&	2204.10600	&	no	&
66	&	2201.12364	&	no	&
90	&	2102.09920	&	no	\\
19	&	2206.00025	&	no	&
43	&	2204.08598	&	no	&
67	&	2202.04671	&	no	&
91	&	2106.05228	&	no	\\
20	&	2205.14161	&	no	&
44	&	2204.09070	&	no	&
68	&	2201.02252	&	no	&
92	&	2106.07659	&	no	\\
21	&	2205.12922	&	no	&
45	&	2201.11131	&	no	&
69	&	2111.01165	&	no	&
93	&	2012.10797	&	no	\\
22	&	2205.11679	&	no	&
46	&	2204.06203	&	no	&
70	&	2111.04203	&	no	&
94	&	2110.14659	&	no	\\
23	&	2205.11318	&	no	&
47	&	2204.06004	&	no	&
71	&	2201.08407	&	no	&&&
\\
24	&	2205.11536	&	no	&
48	&	2204.00847	&	no	&
72	&	2201.05359	&	no	&&&
\\\hline\hline
\end{tabular}
\label{table:papers}
\end{table}

\bibliography{Joycebib}{}
\bibliographystyle{aasjournal}

\end{document}